\documentclass[aps,pra,twocolumn,groupedaddress,floatfix,longbibliography]{revtex4-1}
\usepackage{blindtext,amsmath}
\usepackage{graphicx,enumerate}
\usepackage{float}
\usepackage[export]{adjustbox}
\usepackage[caption=false]{subfig}

\newcommand{\gbb}{g_\mathrm{BB}}
\newcommand{\gib}{g_\mathrm{IB}}

\newcommand{\phiq}{{\phi_\mathrm{q}}(x,\tau)}
\newcommand{\phic}{{\phi_\mathrm{c}}(x,\tau)}
\newcommand{\phiqc}{{\phi^*_\mathrm{q}}(x,\tau)}
\newcommand{\phicc}{{\phi^*_\mathrm{c}}(x,\tau)}

\newcommand{\Xc}{{X_{\mathrm{c}}}(\tau)}
\newcommand{\Xq}{{X_{\mathrm{q}}}(\tau)}
\newcommand{\Pc}{{P_{\mathrm{c}}}(\tau)}
\newcommand{\Pq}{{P_{\mathrm{q}}}(\tau)}
\usepackage{color}

\begin{document}
\title{Stochastic-field approach to the quench dynamics of the one-dimensional Bose polaron}
\author{J. Jager}
\author{R. Barnett}
\affiliation{Department of Mathematics, Imperial College London, London SW7 2AZ, United Kingdom}

\begin{abstract}
We consider the dynamics of a quantum impurity after a sudden interaction quench into a
one-dimensional degenerate Bose gas. We use the Keldysh path integral formalism to derive a truncated Wigner like approach that takes the back action of the impurity onto the condensate into account already on the mean-field level and further incorporates thermal and quantum effects up to one-loop accuracy. 
This framework enables us not only to calculate the real space trajectory of the impurity but also the absorption spectrum. We find that quantum corrections and thermal effects play a crucial role for the impurity momentum at weak to intermediate impurity-bath couplings. Furthermore, we see the broadening of the absorption spectrum with increasing temperature.
\end{abstract}

\maketitle

\section{Introduction}

The interaction of a mobile impurity with a surrounding many-body quantum system is one of the most prominent and oldest problems in condensed matter physics. 
The polaron, initially considered by Landau and Pekar \cite{Pekar1946,Landau1933} to describe an impurity electron interacting with the lattice vibrations of a solid is a prototypical scenario to study quasi-particle formation. In more recent years, neutral atoms immersed in a surrounding ultra-cold gas have attracted widespread attention due to great experimental controllability which enable the study of novel exotic regimes of the polaron.  For example, the impurity-bath coupling can be tuned via Feshbach resonances \cite{Chin2010}.
Here, the fundamental principles at work are the same as in the original problem; the impurity forms a polaron through interacting with the collective excitations of the surrounding superfluid.
While the Fermi polaron has been subject to extensive experimental study \cite{Schirotzek2009,Zhang2012,Kohstall2012,Koschorreck2012,Scazza2017,Cetina2015,Cetina2016,Parish2016,MassignanRPP2014}, the Bose polaron has only been realised in a few experiment  \cite{Catani2012,Jorgensen,Hu,Yan2019,Skou2021}. These experiments hint towards a delicate interplay between equilibrium and 
out-of-equilibrium effects. 

There has been extensive work addressing the Bose polaron in equilibrium \cite{Rath2013,Tempere,Casteels2011,Grusdt2017b,Casteels-PRA2014,Christensen-PRL2015,Ichmoukhamedov2019,Camacho-Guardian2018}. In this work, we focus, in contrast, on the out-of-equilibrium Bose polaron. More precisely, we consider quench dynamics,
involving the abrupt immersion of an impurity into a homogeneous Bose gas,
at finite temperature. 
The quench can be realised through a Feshbach resonant  radiofrequency pulse \cite{Chin2010} that switches on the impurity-condensate interaction
nearly instantaneously.
Such quench dynamics have been studied either at zero temperatures, or focused on the (extended) Fr\"ohlich models (or very similar approximations) \cite{Drescher2019,Shchadilova2016,Volosniev2015,Nielsen2019,Lampo2018, Dzsotjan2020,Mistakidis2019a,Mistakidis2019,Boyanovsky2019,Drescher2020,Ardila2021,Guenther2018,Lausch2018,Charalambous2019}. In much of the equilibrium and some of the out-of-equilibrium treatments, the Lee-Low-Pines transformation \cite{Lee1953}, which eliminates the impurity from the problem at the expense of adding an additional quartic vertex, has proven extremely useful. However, when considering finite temperature, this requires further approximations. 

The vast majority of the existing literature has focused on three-dimensional systems, where the (extended) Fr\"ohlich model is widely applicable. In 1D, the situation is different. In \cite{Grusdt2017b} it was shown that the applicability of Fr\"ohlich-type approximations are somewhat limited in 1D  and while the mass balanced case is integrable for the Fermi-gas \cite{Gamayun2018} no such limit exist in the case of a Bose gas.
Instead, it is natural to
incorporate the effects of the impurity on the condensate already at the mean-field level. This can be done efficiently for a single impurity at zero temperature using the LLP transformation, but this method does not extend to several impurities and is also not straightforward to generalise to finite temperatures. To circumvent those complications, several approaches do not eliminate the impurity from the problem and can address finite temperature \cite{Feynman1955,Lampo2018}, and for example, treat the impurity in a manner related to the coherent state representation of the impurity \cite{Blinova2013}. It has been shown in \cite{Jager2020,Pastukhov2018,Mistakidis2019} that a product wave function within the tree-level approximation, yields inconsistent results with those obtained in the more accurate LLP-frame in 1D. Therefore, it is perhaps more appropriate to treat the impurity in a position-momentum path-integral as it highlights the particle nature of the impurity. Using the coherent state path integral for the condensate has been shown to yield good results in 1D not only for the polaron but also the bipolaron problem \cite{Will2021,Panochko2019,Mistakidis2019}. This is conceptually close to the approach originally developed by Feynman \cite{Feynman1955} and applied to the Bose polaron in \cite{Tempere,Ichmoukhamedov2019} with the main difference being that we do not expand the condensate around a homogeneous density, and our focus lies on out-of-equilibrium phenomena.

In this work,
we develop a conceptually simple and numerically tractable approach to address quench dynamics at finite temperature in a general manner. Ultimately, this is achieved by mapping the dynamics to a set of deterministic differential equations with stochastic initial conditions. By averaging over the different trajectories, expectation values can be calculated within one-loop accuracy. We then use this methodology to study an impurity's evolution after a sudden interaction quench into the bath. We find that the impurity delocalises quickly for weak impurity-bath couplings and that observables like the impurity's velocity crucially depend on the incorporation of quantum and thermal effects. In the case of strong impurity-bath couplings, we observe self-trapping and quantum corrections and thermal effects to be e considerably less pronounced.

This work is organised as follows. In Section \ref{section:Keldysh} we derive the truncated Wigner approximation from the Keldysh path integral representation of the problem at hand. In this section, we also show how to obtain the absorption in the language of semi-classical dynamics. We proceed by specifying the initial Wigner function in Section \ref{section:quench} and show how to regularise the divergences that arise in the one-dimensional setting. We proceed by briefly outlining the numerical considerations in Section \ref{section:numerical}. To conclude, we discuss the results for an impurity at rest and finite momentum in Section \ref{section:results} and outline further directions in Section \ref{section:outlook}.
\section{Methodology}

\subsection{The Keldysh formalism for the Bose polaron}
\label{section:Keldysh}
In the section, we want to outline the derivation of the equations of motion and discuss the truncated Wigner approximation for the polaron problem. We start by considering the Hamiltonian,
\begin{align}\label{eq:Ham_2}
   \hat{\mathcal{H}} = \frac{1}{2M}\hat{P}^2+\int_x  \Bigg\{ \hat{{\phi}}^{\dagger}(x) \Big[\frac{-\partial_x^2}{2m}+\frac{\gbb}{2}\hat{{\phi}}^{\dagger}(x) \hat{{\phi}}(x) \\ \nonumber  + \gib V\left(x-\hat{X}\right) \Big] \hat{{\phi}}(x) \Bigg\},
\end{align}
where $m$ ($M$) denotes the mass of the bosons (impurity atom), $\hat{\phi}(x)$ is the Bose
field operator, $\gbb$ ($\gib$) is the boson-boson (boson-impurity) interaction strength and
$\hat{X}$ ($\hat{P}$) denotes the position (momentum) operator of the impurity. Moreover, we left the interaction potential between the impurities and the condensate general instead of directly assuming s-wave scattering. We will not employ a delta function here but a smoothed out version of it for numerical reasons, that will be discussed later.
In the following, we are going to apply the Kedlysh formalism to (\ref{eq:Ham_2}). The expectation value of an arbitrary observable (see \cite{Polkovnikov2010} for a detailed discussion)  $\Omega(\hat{ X},\hat{ P},\hat{{\phi}}^{\dagger}(x),\hat{{\phi}}(x),t)$ is given by
\vskip-0.1in
\begin{widetext}
\begin{align} \label{eq:expectation}
\langle \Omega(\hat{ X},\hat{ P},\hat{{\phi}}^{\dagger}(x),\hat{{\phi}}(x)),t)\rangle = \int_{X_{0},P_{0},\phi_0(x),\phi_0^*(x)}
W( X_0, P_0,\phi_0(x), \phi_0^*(x)) \int\mathcal{D}[\Xc,\Xq,\\ \nonumber \Pc,\Pq,\phiq,\phic,\phiqc,\phicc] 
\exp\big(iS[\Xc,\Xq,\Pq,\Pc,\\ \nonumber \phiq,\phic,\phiqc,\phicc]\big)\, \Omega_W({ \Xc},{ \Pc},{{\phicc}},{{\phic}},t),
\end{align}
\end{widetext}
where $W$ is the Wigner function that depends on the initial density matrix and will be specified below. $\Omega_W({ \Xc},{ \Pc},{{\phicc}},{{\phic}},t)$ is the Weyl ordered operator of the observable one wants to calculate (again see \cite{Polkovnikov2010} for more details). The subscript $c$ denotes the classical field and the subscript $q$ the quantum field, which describes the quantum fluctuation around the classical saddle point solution. Those two fields arise when mixing forward and backward contour in the Keldysh formalism. The $\mathcal{D}$ denotes the integration over all field configurations in space and time. In contrast to that the first integral in (\ref{eq:expectation}) can be understood as a normal integral in $X_0$ and $P_0$.
The one-loop approximation is now to drop all terms of order two and higher in $\hbar$, which corresponds to an expansion up to second order in the quantum fields. We then find that there are, in fact, only terms linear in the quantum fields, and the action is given by
\begin{align} 
    S = \int_\tau  \Bigg\{  \Big[-\Xq \frac{\rm d}{\rm d \tau} \Pc +\Pq \frac{\rm d}{\rm d \tau} \Xc  
    \\ \nonumber -\frac{\Pq\Pc}{M}\Big] - \int_x \Big[ ( \phicc \big[ -i\partial_\tau -\frac{\partial_x^2}{2m}  \\ \nonumber +\gbb |\phic|^2 + \gib V(\Xc-x)\big] \phiq  \\ \nonumber+c.c.) +\gib\frac{\rm d}{\rm dx} V(x-\Xc)\Xq|\phic|^2 \Big] \Bigg\}   .
\end{align}     
It is now easy to see that all the quantum fields can be easily integrated out and yield functional delta distributions enforcing the classical equations of motion
\begin{align}
    \frac{\rm d}{\rm d \tau}\Xc &= \frac{\Pc}{M} \\
    \frac{\rm d}{\rm d \tau}\Pc &= \gib\int_x \frac{\rm d}{\rm dx} V(x-\Xc)|\phic|^2\\
     i\partial_\tau \phic &= \Bigg(\frac{-\partial_x^2}{2m} +\gbb|\phic|^2\\ \nonumber &\quad + \gib V\left(\Xc-x)\right) \Bigg)\phic.
\end{align}
The only challenge remaining at this point is to integrate over the initial conditions weighted by the Wigner function. In practice this is done by sampling initial conditions according to the Wigner function, solving the classical equations of motion and then averaging the desired observable over the calculated trajectories. In this framework it is now straightforward to calculate the impurity dynamics. Before proceeding we would like to make some remarks about the classical equations of motion and make contact
with the equilibrium case to show that even without taking the first-order correction into account those equations give satisfactory results in the limiting { case of heavy impurities}. For the equilibrium case we assume the impurity to travel at constant velocity $\frac{\rm d}{\rm d \tau}\Xc = v$,  implying $\frac{\rm d}{\rm d \tau}\Pc =0 $, which in turn tells us that $|\phic|^2$ is symmetric around the impurity position. Together with the equilibrium assumption this directly leads to the conclusion that the bosonic field takes the following form $\phic = \phi_c(x-v\tau)=\phi_c(x-\Xc) $. In the equilibrium setting we consequently find $i\partial_\tau \phic = -iv\partial_x \phi_c(x-\Xc) $. Putting it all together and defining $\tilde{x} = x-\Xc$ we find the equilibrium equation
\begin{equation}
   \left(\frac{-\partial_{\tilde{x}^2}}{2m} +\gbb|\phi_c(\tilde{x})|^2 +iv\partial_{\tilde{x}} + \gib V(\tilde{x}) \right)\phi_c(\tilde{x}) = 0.
\end{equation}
We can now compare this equation with the one obtained by performing a Lee-Low-Pines transformation and find that it has in fact the same form as the one found in \cite{Jager2020,Mistakidis2019,Volosniev2017,Panochko2019}, where it has been shown that 
 quantities obtained like the effective polaron mass or the polaron energy are in excellent agreement with results obtained by quasi-exact quantum monte carlo methods. The only difference is that instead of the reduced mass the boson mass appears in front of $\partial_x^2$, which can be traced  back to the fact that in this derivation the effect of normal ordering is lost. 
However, this effect is unimportant for heavy impurities. To summarize, we showed that the equation obtained by employing a coherent state ansatz for the bosonic field paired with a position momentum representation for the impurity reduce to the correct mean-field equations in the equilibrium case if the impurity mass is large.

 Another quantity of great interest is the impurity Green's function \cite{Dzsotjan2020,Rath2013,Shchadilova2016} from which the absorption spectrum can be calculated by taking the Fourier transformation. The impurity Green's function is defined by
\begin{align}
G(t) = \mathrm{Tr}\left\{ \exp\left(i\hat{H_0}t\right)\hat{\rho} \exp\left(-i\hat{H}t\right)\right\}
\end{align}
where $\hat{H}_0$ stands for (\ref{eq:Ham_2}) with $V(x- \hat{X}) = 0$ and $\hat{\rho}$ is the initial density matrix. We now observe that this has the same structure as the trace that is considered to derive the Keldysh path integral, with the only difference, that the forward and backward contour differ by an extra interaction term. We can therefore perform the same steps as when deriving (\ref{eq:expectation}). The only difference in $S$ is the resulting impurity boson interaction 
\begin{align} \label{eqn:S_int}
    S_\mathrm{int} = \int_{\tau,x}\Bigg[ \big( \phicc\big[\frac{1}{2} \gib V^{(n)}(\tilde{x}) \big] \phiq  \\ \nonumber+c.c.\big)+\Xq\frac{\gib}{2} V^{(1)}(\tilde{x}) |\phic|^2\\ \nonumber\Xq + \gib V(\tilde{x})(|\phic|^2\\ \nonumber+\frac{|\phiq|^2}{4})  +\frac{\gib}{8}|\phic|^2 V^{(2)}(\tilde{x})\Xq^2  \Bigg],
\end{align}
 where we have introduced the notation $\frac{\rm d^n}{\rm dx^n}V(x-\Xc)=V^{(n)}(\tilde{x})$
The resulting magnitude of the interaction is changed by a factor of $1/2$ and a new purely classical term arises . Lastly, we note that there is also a quadratic term in $\phiq$ and $\Xq$ now. If we want to keep the accuracy up to one loop order this term has to be considered, which somewhat complicates matters. An ad hoc approximation is to drop this term altogether and therefore staying in a strictly semi-classical regime.
To see when this approximation is justified one can simply compare the arsing terms and their order of magnitude. We note, that the $|\phiq|^2$ term directly competes with the $|\phic|^2$ term. As long as one is within the applicability region of a general c-field treatment, $|\phiq|^2$ will be small compared with $|\phic|^2$, whenever the condensate deformation is not large, which corresponds to small and intermediate $\eta$. For the corrections in the impurity degrees of freedom one has to compare $\frac{\gib}{2} V^{1}(\tilde{x})|\phic|^2$ with $V^{(2)}(\tilde{x})\Xq^2|\phic|^2$. We note that through partial integration the derivative terms can be brought onto bosonic field variables.  One then realizes, that the second derivative of the fields is going to be small compared to the first derivative for weak coupling. Additionally,  for large impurity masses $M$ the magnitude of $X_q$ will stay small. Those two considerations show that one would expect the absorption spectrum  yield reliable results for weak to intermediate impurity-boson coupling and potentially a wider range of couplings if the impurity is sufficiently heavy. We refer to section \ref{sec:validity} for more details on the validity the approach presented here.

We also note that this approximation is only made when calculating the absorption spectrum and the impurity Green's function and all dynamical results do not rely on this approximation.  Henceforth, the additional stochastic term will be dropped.
This leaves us with the expression
\begin{equation}
    G(t) = \langle\exp\left(i\int_{\tau,x}  \gib  V(\Xc-x)|\phic|^2\right)\rangle_W,
\end{equation}
where we denote the average with respect to the initial Wigner function as $\langle...\rangle_W$. We are now in the position to calculate the real space trajectories of the impurities for finite temperature as well as the absorption spectrum. 

\subsection{The quench protocol and the initial Wigner function}
\label{section:quench}
In the following, we want to specify the quench protocol and the initial Wigner function. We start by briefly outlining the initial state and then specify the Wigner function for a 1D quasi condensate. Here, we will also discuss all the regularisation necessary to arrive at a divergence-free quasi condensate description.
The quench protocol is the following: we start with a free impurity and an interacting superfluid at temperature $T$. At $t=0$, the interaction between the superfluid and the impurity is turned on instantly. Experimentally this is realised through a Feshbach resonance \cite{Chin2010}. 
We assume that the initial density matrix can be written as a direct product of the state of the impurity (which is assumed to be pure) and the thermal density matrix of the superfluid 
\begin{align}
    \hat{\rho} = \hat{\rho}_\phi \otimes |\psi\rangle\langle\psi|,
\end{align}
As a consequence, the Wigner function also factorises, and we can sample the initial conditions independently. For the condensate, we employ a quasi condensate description; in 1D, this is best done by employing a density and phase representation. We note that this has been used before in \cite{Martin2010}, in the trapped gas case. Since we want to focus on a homogeneous gas in continuum here, we need to regularise the non-condensed part. 
In this representation, the condensate field operator can be written as
\begin{align}
    \hat{\phi}(x,0) = \sqrt{n_0 + \delta\hat{\rho}(x)}\exp(i\hat\theta(x)).
\end{align}
The density operator and the phase operator can be expressed within the Bogoliubov approximation \cite{pitaevskii} as
\begin{align}
    \hat{\theta}(x) &= \frac{-i}{2\sqrt{\rho_0}}\sum_k \Big[(u_k+v_k)e^{ikx}\hat{a}_k -\mathrm{h.c.} \Big]\\
    \delta\hat{\rho}(x) &= \sqrt{n_0}\sum_k \Big[(u_k-v_k)e^{ikx}\hat{a}_k +\mathrm{h.c.} \Big],
\end{align}
where $u_k$ and $v_k$ are the usual Bogoliubov modes. 
For this treatment to be valid one has to be in the vicinity of a weakly interacting Bose gas, which can be characterised by the Tonks parameter $\gamma=1/(2 n_0^2 \xi^2)$ \cite{Girardeau1960,Lieb1963a}
which should be less than unity 
where $\xi = \frac{1}{\sqrt{2\gbb n_0}}$ is the healing length.   We refer to the section \ref{sec:validity} for a detailed discussion of the validity of the presented approach. In the path integral this corresponds to a shift of variables meaning that instead of integrating over $\phi^{(*)}_0(x)$, we integrate over $a^{(*)}_k$, corresponding to the operators $\hat{a}^\dagger_k$. In the standard way, we can now write down the thermal Wigner function (within the coherent state representation) for the $a_k$ \cite{Polkovnikov2010}
\begin{align}
    W(a^*_k,a_k) = \frac{2}{\pi}\tanh\bigg(\frac{\omega_k}{2T}\bigg)\exp\Big\{-2|a_k|^2\ \tanh\bigg(\frac{\omega_k}{2T}\bigg)\Big\},
\end{align}
with the Bogoliubov dispersion $\omega_k$. From this, it can be seen immediately, that the average condensate particle number is given by $N_0 /L = \rho_0$. In order to  account for the quantum and thermal depletion, we fix the total particle number $N$ and then choose $N_0$ according to $N_0 =N-N_d$ (this is done for every realisation), where after proper regularisation (see \cite{Salasnich2016,Stoof-book})
\begin{align}\label{eq:part_number}
    N_d = \sum_k \Big[\frac{e_k-\omega_k}{2\omega_k}+\frac{\mu}{2e_k+2\mu} +\frac{e_k}{\omega_k}\big( a^*_k a_k -\frac{1}{2} \big)\Big],
\end{align}
with the single-particle dispersion $e_k = k^2/(2m)$. Here, a first-order T-matrix approximation was employed, and in line with the Bogoliubov theory up to one loop, $\mu$ is the chemical potential within the Bogoliubov approximation and hence is not temperature-dependent. The $1/2$ is needed to cancel the extra factor that stems from the symmetric ordering. After averaging, this reproduces the expected result for a thermal quasi condensate in 1D. It should be noted that $\mu$ has to be chosen consistently with the total particle number, which can be done by fixing one reference point, where the total particle number is known. Henceforth we assume a mean-field density at $T=0$. This then fixes $\mu$ in the Bogoliubov approximation through $\mu = \gbb n_0(T=0)$  We can now use (\ref{eq:part_number}) to determine the total particle number which remains fixed throughout the calculation. We can now sample individual realisations of the condensate, whose description is free of IR and UV divergences. Upon closer inspection, one might realise that even though the mean of the phase and density corrections are zero, the variance scales up to linearly with the system size $L$. A direct result of this is that the computational time needed to achieve convergence also scales with the system size. This computational challenge can be tackled by increasing the system size gradually until the results are independent of the system size and then validating certain data points with larger system size. Furthermore, this restricts the discretisation of space, as outlined in \cite{Castin}, which will be discussed in the next section. Because the effect of the impurity is local, we find relatively low dependence on the system size already for small systems. 
Lastly, we assume that the impurities are not entangled (namely can be represented by a product wave function) and are localised in space around $x_0$ or equivalently in momentum space localised around $q$ at $t=0$. It is therefore natural to choose a wave packet as their initial wave function
\begin{align}
    \langle x | \psi \rangle = \frac{1}{\pi^{1/4} \sqrt{a_0}} \exp\big(-(x-x_0)^2/(4a_0^2)\big).
\end{align}
Here $a_0$ is an external parameter that determines how localised the initial state is. The Wigner function in this setting is well known to be \cite{Polkovnikov2010}
\begin{align}
    W(X_{0},P_{0}) = 2\exp \left(-2a_0^2(P_{0}-q)^2-(X_{0}-x_0)^2/(2a_0^2) \right).
\end{align}

\subsection{Numerical considerations}
\label{section:numerical}
In this section, we will show that three quantities can describe the whole parameter space, and we will briefly outline the discretisation of space and all the subtleties involved.
We now define the following scaled parameters
\begin{align}
    \tilde{\tau} = \tau/\tau_s, \quad \tilde{x}= x/x_s, \quad \tilde{\phi}(\tilde{x}) = \sqrt{x_s}\phi(x),\\ \nonumber \quad \tilde{P} =  P x_s, \quad \tilde{X} =  X/x_s,
\end{align}
where we choose $\tau_s = \tilde{\xi}/c$, $x_s =\sqrt{2}\xi=\tilde{\xi}$, and where the speed of sound is given by $c=\sqrt{\frac{\gbb n_0}{m}}$. 
Dropping the twiddle we then find the Hamiltonian
\begin{align}\label{eq:Ham_dedim}
   \hat{\mathcal{H}} = \alpha   \hat{ P}^2+\int_x  \Big\{ \hat{{\phi}}^{\dagger}(x) \big[\frac{-\partial_x^2}{2}+\frac{\sqrt{\gamma}}{2}\hat{{\phi}}^{\dagger}(x) \hat{{\phi}}(x)  \\ \nonumber +   \eta \sqrt{\gamma} V(x-\hat{ X)} \big] \hat{{\phi}}(x) \Big\},
\end{align}
where $\eta$ remains unchanged and $\alpha = \frac{m}{M}$.
From here the following equations of motion can be obtained
\begin{align}
    &\frac{\rm d}{\rm d \tau}\Xc = \alpha \Pc \label{eq:eom_x}\\
    &\frac{\rm d}{\rm d \tau}\Pc = \eta\sqrt{\gamma}\int_x \frac{\rm d}{\rm dx} V(x-\Xc)|\phic|^2\\
     &i\partial_\tau \phic = \Big(\frac{-\partial_x^2}{2}   +\sqrt{\gamma}|\phic|^2  \\ \nonumber & \quad +  \eta \sqrt{\gamma} V(\Xc-x)\Big)\phic.
\end{align}
In order to tame the extensive variance and ensure numerical stability, we have to choose the discretisation $l = L/N_\mathrm{grid}$ as outlined in \cite{Castin}, namely $l$ has to be large enough to satisfy $\rho_0 l \gg 1$, while at the same time ensuring that the energy cut-off introduced by $l$ does not alter the physics. This translates to $l < \xi,\lambda$, where $\xi$ is the healing length that sets a natural length scale for our problem and $\lambda$ is the thermal de Broglie wavelength. Lastly, we note that we make the following choice for the interaction potential
\begin{equation}
    V(x) = \frac{1}{\sqrt{2\pi l^2 }}e^{-x^2/(2l)},
\end{equation}
which converges to the delta distribution as $l\rightarrow 0$, but has the advantage of being smoother than $\delta_{x}/l$, where $\delta_x$ is the Kronecker delta.

\subsection{Validity of  formalism} \label{sec:validity}

In this subsection we will address the regime of validity of the formalism used in this work. We start by giving some general arguments for the validity of the approach followed by a term-by-term discussion of the higher-order corrections and their order of magnitude. While for the absorption spectrum we have to restrict our considerations to weak or moderate boson-impurity couplings, we would like to stress that for all dynamical properties calculated (excluding the absorption spectrum) the result is strictly non perturbative in $\gbb$ and $\gib$.  Hence we can safely say that the method presented is valid for $\gamma \leq 1$, which is equivalent to the range for the Gross-Pitaevskii  equation, which corresponds to the tree level approximation of (\ref{eq:Ham_2}). The same reasoning applies to $\gib$, leading to the conclusion that our results, at least for short times are valid across the whole range from weak to strong impurity-boson couplings. For longer times one has to consider the deformation of the condensate. Another interesting point to consider is that as soon as $\alpha$ is small (meaning the impurity is heavy) the accuracy of the presented approach will improve further, since the impurity will behave more classically. In fact $\alpha$ can be seen as a strict control parameter for the corrections arising due to higher order terms in $\Xq$. Combining those considerations with the known fact that in general the truncated Wigner approximation is exact for short time scales \cite{Polkovnikov2010}, we can conclude that for short to intermediate time scales our results are trustworthy for all values of boson-boson and boson-impurity coupling. For  weak coupling it is ab initio reasonable to assume that the presented results hold for longer time scales, since higher order quantum corrections should accumulate slowly if at all.  
However, no such universal statement can be made for strong couplings. 
There are two contributions of order $\hbar^2$ or higher, which were neglected in our approach and would have to be added to (\ref{eqn:S_int}) if one wanted to solve the problem exactly, the first one coming from the boson-boson interaction and the second one being due to the impurity-boson interaction. The first one takes the form
\begin{equation}
    \sqrt{\gamma}\left( |\phi_q(x,\tau)|^2\phi_q(x,\tau)\phi^*_c(x,\tau) +{\rm c.c.} \right).
\end{equation}
We note that this term is closely related to the standard Bogoliubov approximation, with the main difference being that the classical field here is taken to be the deformed field. We note that in no point of our simulations the expectation value of $\langle |\phi^*_c(x,\tau)|^2 \rangle$ falls  below the value of $3\tilde{\xi}$, meaning that the it is safe, in the spirit of the 
well-established Bogoliubov approximation, to neglect higher-order terms in the quantum fields, which only scale linearly in $\phi_c$ and are of $\mathcal{O}(\phi_c\phi^3_q)$ which is certainly small compared to $\mathcal{O}(|\phi_c(x,\tau)|^3\phi_q)$, as long as the density of the condensate stays larger than the healing length. We note that in the case of very low density gases and strong boson-impurity coupling the above argument comes under question and it is a priori not clear whether the outlined method is reliable in this regime.  
This regime was not investigated in the present work. This argument is further underlined by the remarkable accuracy of pure c-field methods (which do not take first-order quantum corrections into account) for the equilibrium polaron, see \cite{Jager2020, Drescher2020,Volosniev2017,Will2021}, where the c-field approximation was shown to also hold for low-density gases and strong coupling, and the fact that the boson-boson interaction to all orders in $\gbb$ and to first order in $\hbar$ are already taken into account in our calculations. A somewhat more complicated expression is obtained when considering the impurity-boson interaction term. 
Here, one finds that already in (\ref{eqn:S_int}) all orders of $\Xq$ are present and the higher order terms take the following form
\begin{align}
    \eta \sqrt{\gamma} \Bigg( -\frac{|\phiq|^2}{4}\sum^\infty_{n=0} V^{(2n+1)}(\tilde{x})\frac{X^{(2n+1)}_q}{4^n(2n+1)!}\\ \nonumber
    -|\phic|^2\sum^\infty_{n=1} V^{(2n+1)}(\tilde{x})\frac{X^{2n+1}_q}{4^n(2n+1)!} \\ \nonumber
    +(\phicc\phiq+c.c.)\sum^\infty_{n=1} V^{(2n)}(\tilde{x})\frac{X^{2n}_q}{4^{n-1}(2n)!}  \Bigg).
\end{align}
Again for approximating the impact of those terms it is convenient to bring the derivative terms to the bosonic field operators. 
It then becomes clear that all corrections can be understood as a gradient expansion in the bosonic field operators, whose impact is certainly small for short time scales and also for longer time scales as long as the coupling stays moderate. Besides the gradient terms we also note that each correction is accompanied by an increasing power of $\Xq$ terms, which are also small at short times and whose impact can be controlled by $\alpha$.

To summarize, while the validity of this approach for very large couplings and long time scales can not be judged a priori, we note that for short time scales the results here hold regardless of impurity-boson coupling strength and also note that 
$\alpha$ serves as a control parameter for the approximation in the impurity degrees of freedom.
\begin{figure}
    \includegraphics[scale = 1.0]{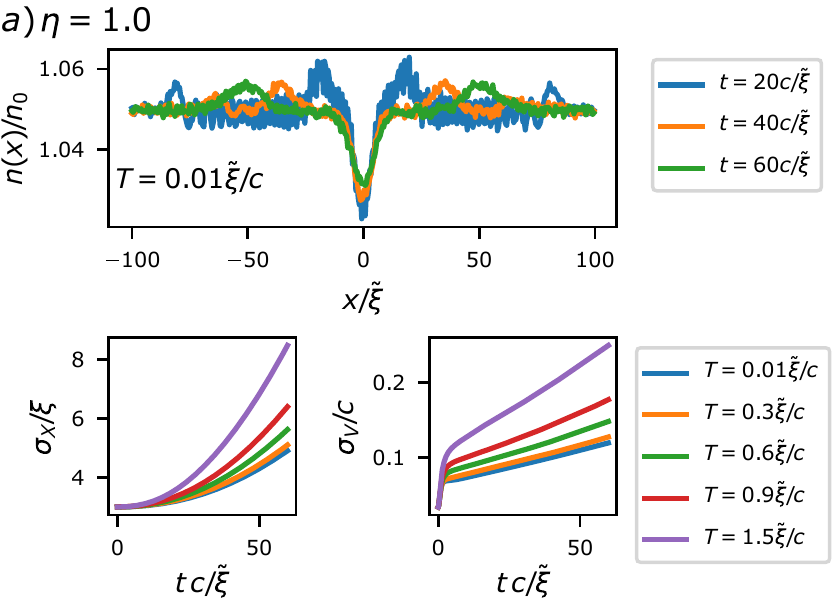}
    \includegraphics[scale = 1.0]{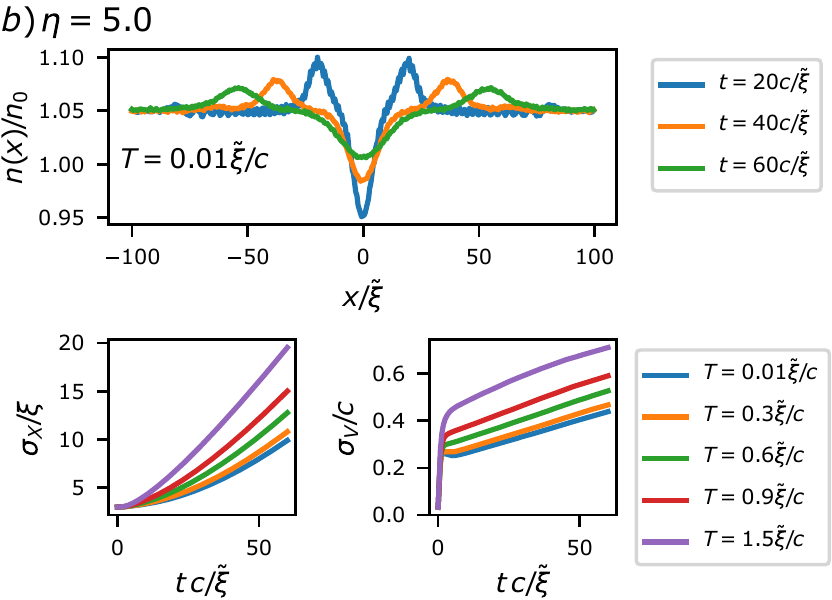}
    \includegraphics[scale = 1.0]{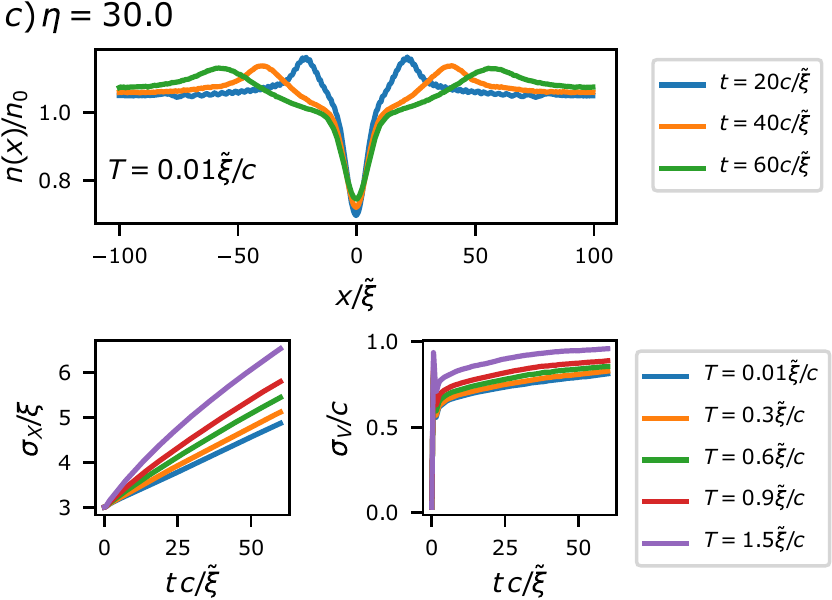}
  \caption{The superfluid density, the position variance and the variance of the velocity for different repulsive interactions and temperatures. In all plots the parameters are $\alpha =0.2$, $\gamma =0.04$, $n_0 = 5\tilde{\xi}$ and $a_0 = 3/\tilde{\xi}$. As expected, we can clearly see that the impurity deforms the condensate over time. It also becomes apparent that the impurity delocalises over time and that this effect is slowed down with increasing $\eta$, which can be understood by a self-trapping argument given in the main text.  }
  \label{fig:RMS}
\end{figure}
\section{Results}
\label{section:results}
\subsection{Post-quench density profile}
\begin{figure}
\includegraphics[scale = 1.0]{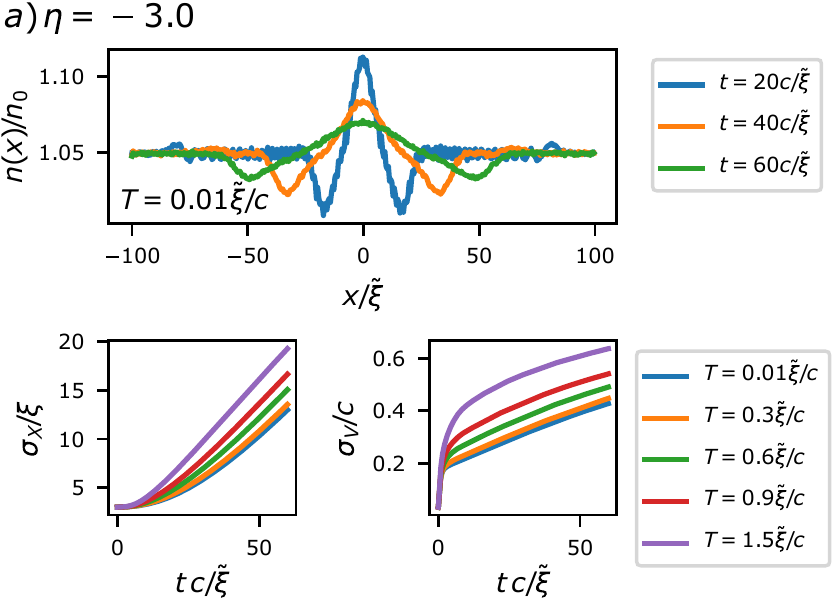}
\includegraphics[scale = 1.0]{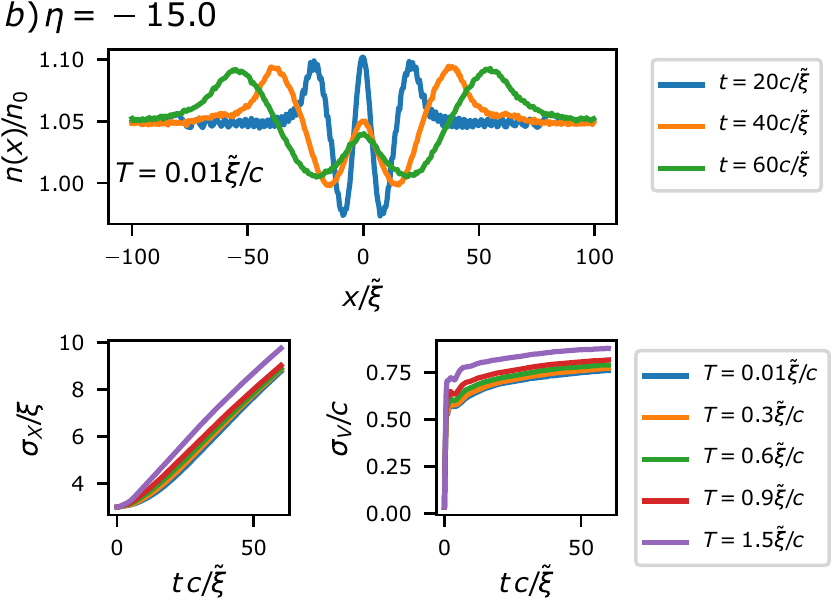}
  \caption{The superfluid density and position variance and velocity variance for different attractive interactions and temperatures. In all plots the parameters are as follows  $\alpha =0.2$, $\gamma =0.04$, $n_0 = 5\tilde{\xi}$ and $a_0 = 3/\tilde{\xi}$. We can see that the impurity attracts surrounding particles, which in turn depletes the condensate. As can be seen in (b), this appears as a repulsively interacting polaron.} 
  \label{fig:attrRMS}
\end{figure}
In this subsection, we focus on the density of the condensate for repulsive and attractive impurity-bath couplings at different times after the quench and supplement those findings with the variance of the position $\sigma^2_{X} = \langle (\hat{X} -\langle \hat{X} \rangle)^2 \rangle$ and the variance of the velocity $\sigma^2_{V} = \langle (\hat{P}/M -\langle \hat{P}/M \rangle)^2 \rangle$.
We find a dynamically distinct behaviour for weak and strong couplings on the repulsive and the attractive side.

In Fig. \ref{fig:RMS}, the condensate density at different times and the evolution of the variance of the position and the variance of the velocity for repulsive interactions are shown. 
Before discussing the results in more detail we notice that even for $\eta = 50$ the minimum of the condensate density is still larger than $3\tilde{\xi}$, indicating that the approximations  made later for the absorption spectrum are justified. We note that for weak coupling, the impurity delocalises faster than a free impurity would. For stronger couplings, the impurity stays localised much longer in time, indicating self-trapping. The velocity variance saturates after a finite time, and the time scale is inversely proportional to the impurity boson coupling. 
We note that this can be explained by realising that two competing effects are determining the dynamics. Namely, the impurity tends to distribute the repulsion equally throughout the condensate, causing the impurity to delocalise and the opposing effect of self-trapping, where the impurity deforms the condensate and then self-traps in the deformation. It is intuitively clear that self-trapping will not occur for weak couplings, which explains the different behaviours seen in Fig. \ref{fig:RMS}. It can also be seen that the variance of the impurity velocity can exceed the speed of sound, 
which is associated with the emission of non zero energy excitation, indicating an energy transfer from the impurity to the bath, which has also been observed in \cite{Mistakidis2019b}.
The temperature influences the value with which the position variance and velocity variance saturate; it does not significantly influence the timescale. For strong coupling, the temperature dependence becomes relatively weak, which can be explained by noting that the impurity-boson scattering length determines the relevant energy scale, which is much larger than the thermal length in this case.

In Fig \ref{fig:attrRMS} the same situation for attractive couplings is shown. Here the difference between strongly attractive and moderate attractive couplings becomes obvious. We observe that in the case of moderately attractive couplings, the impurity not only diffuses but also forms a purely attractive polaron. In contrast, for strongly attractive couplings, an attractive polaron with repulsive interactions is observed, and the time scales of the polaron formation are prolonged. This difference is a dynamical effect and can be understood by noting that when the interaction is turned on, particles from the condensate start to accumulate around the impurity, which in turn depletes the condensate around the impurity. The superfluid is interacting itself, and therefore the depletion is filled by the particles around it, with the time scale being set by the boson-boson interaction. Meanwhile, the impurity-boson interaction strength determines the number of particles that can accumulate around the impurity before the boson-boson action prevents further accumulation. At the same time the impurity delocalises, which prevents the formation of a well-defined peak around the impurity and the interaction can thus seem repulsive for a long time scale. Ultimately this is of course only a metastable state. This process continues for a longer time when the impurity-boson interaction is large, resulting in a polaron that looks repulsive which can be observed in Fig \ref{fig:attrRMS}(b).

\subsection{Impurity velocity}
Another quantity that is of great interest is the impurity velocity. The out-of-equilibrium case gives insight into the polaron formation and the time scales at work. It is also of great importance for the equilibrium case since it can be used to calculate the effective mass of the polaron \cite{Grusdt2015}.

\begin{figure}
	\centering
	\includegraphics[scale = 1.0]{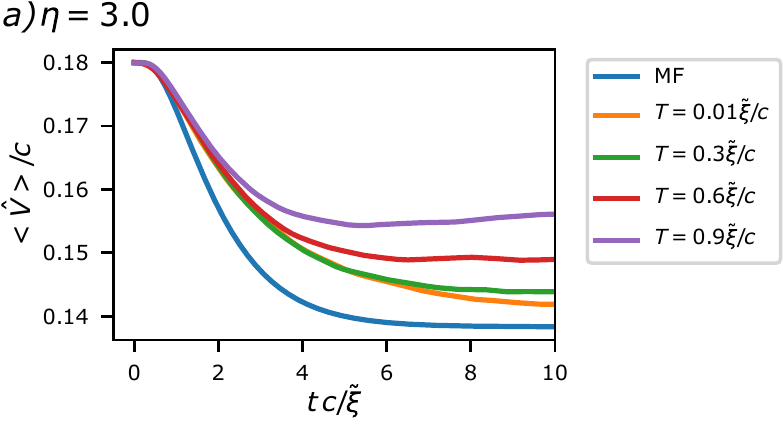}
	\includegraphics[scale = 1.0]{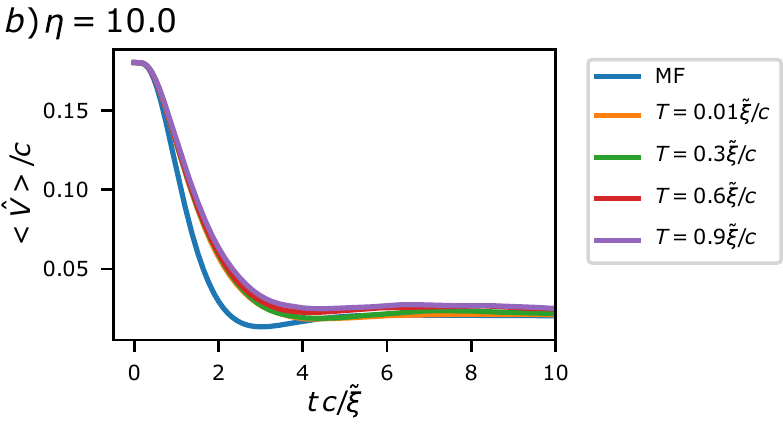}
	\caption{The impurity velocity over time for different temperatures. The parameters are 
	$\alpha =0.2$, $\gamma =0.04$, $n_0 = 5\tilde{\xi}$ and $a_0 = 3/\tilde{\xi}$. We note that the quantum corrections have a big influence for weak impurity-bath interactions.}
	\label{fig:velocity}
\end{figure}

Here, the impurity is not at rest when the quench occurs but instead carries some finite momentum. The sudden quench of the impurity-boson interaction leads to a momentum transfer from the impurity to the surrounding particles, and we expect a slow down of the impurity. The time evolution for the velocity of the impurity is depicted in Fig. \ref{fig:velocity}. Here it becomes apparent that quantum corrections have a noticeable impact on the evolution of the velocity at weak to intermediate coupling. This can easily be understood by noting that the impurity is treated as a point particle on the mean-field level, and as observed in Fig \ref{fig:RMS}, which is not a valid approximation for weak couplings. This also explains why the MF velocity is lower than the corrected solution. We also note that the steady-state velocity increases with the temperature, which can be understood by noting that the surrounding gas has a higher average squared velocity for increasing temperature, and therefore the momentum transfer will be smaller. In contrast, for strong couplings, the impurity stays localised and approximating it as a point particle is, therefore, less of a simplification. The same behaviour can be observed for the temperature dependence, which is more critical for weak and intermediate couplings, which is again due to the scattering length being larger than the thermal length.
We also note that the impurity transfers some of its momentum to the Bose-gas and then relatively quickly reaches an equilibrium velocity for not too strong interactions. For stronger interactions, we observe different behaviour. After an initial abrupt slow down of the impurity, the impurity velocity changes sign, which directly results from the back action from the condensate. We note that a similar abrupt slowdown has been observed in the three dimensional setting in \cite{Grusdt2018a}. The velocity then performs a damped oscillation around its final velocity. 
\subsection{The absorption spectrum}
\begin{figure}
\includegraphics[scale = 0.98]{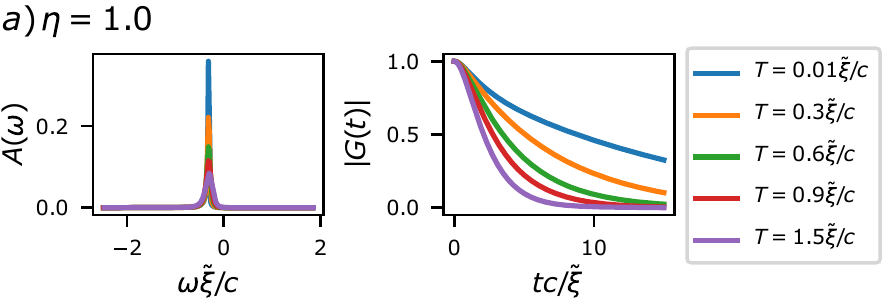}
\includegraphics[scale = 0.98]{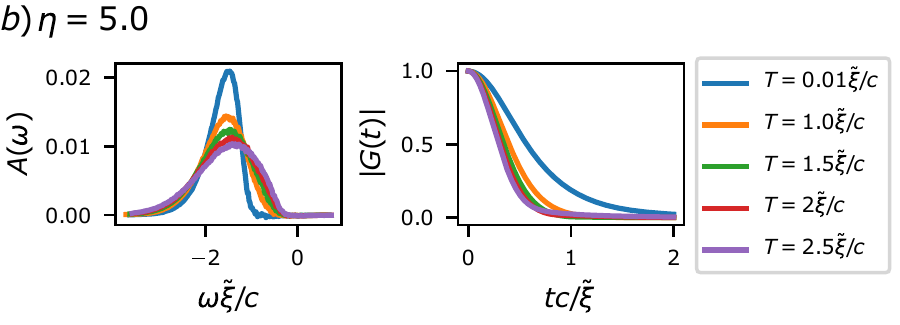}
  \caption{The absorption spectrum $A(\omega) = 2 \mathrm{Re}\int^{\infty}_0 G(t) e^{i\omega t} \, \mathrm{d} t$ for different temperatures $T$ calculated using the truncated Wigner approximation for $\alpha =0.2$, $\gamma =0.04$, $n_0 = 5\tilde{\xi}$ and $a_0 = 3/\tilde{\xi}$} 

  \label{fig:absorb}
\end{figure}
Next we turn our attention to the (injection) absorption spectrum $A(\omega) = 2  \mathrm{Re} \int^{\infty}_0 G(t) e^{i\omega t} \, \mathrm{d} t$ \cite{Cetina2016,MassignanRPP2014}, which can be measured using Ramsey spectroscopy \cite{Jorgensen,Cetina2015,Cetina2016}. It can be calculated by taking the Fourier transform of the impurity green's function, which characterises the dephasing of the system and is closely related to the Lochschmidt echo \cite{Vanicek2016}.
The absorption spectrum gives essential information about the polaron formation and can be used to estimate the polaron energy and lifetime \cite{Kain2014,Mahan2000}.
At this point, we also want to stress that pure mean-field (MF) calculations in position momentum basis are not sufficient to calculate the impurity Green's function. This can be seen by noting that there is no averaging for a classical calculation, which means there is no dephasing between different trajectories, and therefore one always obtains $|G|=1 $ in purely classical calculations.

Our results are depicted in Fig. \ref{fig:absorb}, it can be observed that the quasi-particle peak widens with increasing temperature, which is also reported in 3D
\cite{Dzsotjan2020}. However, in contrast to the 3D case, \cite{Shchadilova2016}, where the extended Fr\"ohlich model was considered, we do not find several peaks on the repulsive side. We also note that the overall amplitude decreases with $\eta$, and the quasi-particle peak gets washed out with increasing $\eta$, which is a direct consequence of the orthogonality catastrophe
\cite{Fogarty2020,Guenther2020,Campbell2014}, hence the emergence of a clear quasi-particle peak comes under question. We note that the absorption spectrum shows a functional dependence associated with quasi-particle behaviour, and we do not find an infrared dominated regime as observed in other one dimensional systems \cite{Kain2014}. Those findings are supplemented by the overlap $G(t)$. Here we can see another vastly different feature of the one-dimensional case, compared to the three-dimensional case \cite{Dzsotjan2020,Shchadilova2016}, namely that $|G|$, approaches zero even for moderate couplings, signalling the onset of the orthogonality catastrophe. As expected, the dephasing becomes more rapid with increased temperature and increased $\gib$.
\section{Summary \& outlook}
\label{section:outlook}
In summary, by leveraging the Keldysh formalism, we derived a truncated Wigner approach to study dynamical properties of the bose polaron in 1D. This allowed us to reduce the problem to simulating semi-classical equations of motion with stochastic initial conditions. We showed how to adequately account for temperature effects of the surrounding bath by sampling the phase and density of the condensate. We also discussed how to regularise the arising divergences that typically occur in such one dimensional systems. 
The method presented here takes the back action of the impurity onto the condensate into account and is therefore applicable from weak to strong impurity bath couplings. 

We then used this framework to calculate the dynamics of an impurity after sudden immersion in a surrounding bath and the absorption spectrum. By considering the condensate density and the position/velocity variance, we showed that there is a distinct dynamical behaviour associated with the strong and weak coupling regime, namely our results indicate self-trapping of the impurity for strongly repulsive interactions, and we also find a repulsive polaron on the attractive side.
We also investigated the temperature dependence of the polaron formation and found a substantial influence of quantum corrections on dynamical properties like the velocity of the impurity, showing the necessity to go beyond pure mean-field considerations. Lastly, we considered the absorption spectrum and the impurity Green's function. Here, we observed a clear quasi-particle peak for weak to intermediate couplings. In contrast to that, we see that the quasi-particle peak is washed out for strong couplings and that temperature effects widen the quasi-particle peak. In contrast to the higher dimensional case, the impurity Green's function approaches zero even for weak couplings.

At this point, we want to stress that our approach is neither limited to 1D nor a single impurity. It could therefore serve as an exciting starting point to explore higher dimensional systems, as well as the interplay of several impurities. 
While the generalization to several impurities is quite straightforward we want to stress that, as pointed out for example in \cite{Ardila2020,Khan2021,Lingua2018}, the generalisation to higher dimensions is highly non-trivial in general. First we note that in higher dimensions it is not possible to use bare contact interactions for the boson-boson interaction and the boson-impurity interaction simultaneously when employing the approach. One has to resort to using more realistic interaction potentials for at least one of them as has, for example, been done in the three dimensional context in \cite{Drescher2020,Guenther2020}. Another major challenge is a purely numerical one,  as it becomes more costly to sample the Bose fields in higher dimensions. 
Nevertheless we expect that the presented method may be paired with some small approximations to address higher dimensional systems.

\section*{Acknowledgments}
We would like to thank Michael Fleischhauer and Martin Will for fruitful discussions.
J.J. \ is grateful  for support from EPSRC under Grant
EP/R513052/1.
R.B. is grateful for support from a Cecilia Tanner Research Impulse Grant.

\bibliography{library.bib} %
\end{document}